\documentclass[twocolumn,aps,prb,floats,showpacs,superscriptaddress]{revtex4}
\usepackage{amsmath}

\newcommand{\be}{\begin{equation}}
\newcommand{\ee}{\end{equation}}

\usepackage{epsfig}
\usepackage{graphicx}% Include figure files
\usepackage{bm}% bold math
\usepackage{amssymb}%

\def\gapp{\lower.35em\hbox{$\stackrel{\textstyle>}{\sim}$}}
\def\lapp{\lower.35em\hbox{$\stackrel{\textstyle<}{\sim}$}}

\begin{document}

\bibliographystyle{apsrev}

\title{Domain Wall Motion in Thin-film Magnets/Topological Insulator Junctions}

\author{Yago Ferreiros}
\email{yago.ferreiros@csic.es}
\author{Alberto Cortijo}
\email{alberto.cortijo@csic.es}

\affiliation{Instituto de Ciencia de Materiales de Madrid, CSIC, Cantoblanco, 28049 Madrid, Spain.}

\begin{abstract}
We derive the equations of motion of a domain wall in a thin-film magnet coupled to the surface states of a topological insulator, in the presence of both an electric field along the domain wall and a magnetic field perpendicular to the junction. We show how the electric field acts as a chirality stabilizer holding off the appearance of Walker breakdown and enhancing the terminal velocity of the wall. We also propose a mechanism to reverse the domain wall chirality in a controllable manner, by tuning the chiral current flowing through the wall. An input from a weak perpendicular magnetic field is required in order to break the reflection symmetry that protects the degeneracy of the chirality vacuum.
\end{abstract}

\maketitle

\section{Introduction}
\label{sec:intro}
Within the field of spintronics, the electric control of magnetic domain wall (DW) motion holds one of the prominent places. It is of considerable interest both from fundamental as from the perspective of applied science, and this is so since the experimental realization of the magnetic ``race-track" technology\cite{PHM08,TMR10}. Traditionally, there were two main ways of acting on a magnetic DW: the application of an external magnetic field and the interaction with an electronic current which is spin-polarized and is hosted by the ferromagnet. As it is well known since the early studies of DWs in the presence of an external magnetic field, the DW motion is severely reduced when the magnetic field exceeds some critical value and the system enters in the so-called region of Walker breakdown\cite{STK11}, caused by DW structural instabilities, setting a limitation for actual high speed operation devices. Trying to stabilize the internal structure of the DW, so higher velocities can be reached, has become an important task \cite{GCC05}. In this context the presence of the spin-torque field, while still suffering from the upper Walker breakdown bound, significantly enlarges the DW velocity \cite{TKS08}. There has been recent experimental success on delaying the appearance of Walker breakdown in the context of current-induced motion, by Rashba field mediated chirality stabilization \cite{ITH11}. Also, recent experimental progress has been done in controlling the DW motion by using external electric fields. The strategies in this case consist on changing the carrier density by field effect and thus controlling the spin-torque effect\cite{OCM00}, or modifying the perpendicular anisotropy  by electric fields\cite{SBF11}.

\begin{figure}
\begin{center}
\includegraphics[angle=0,width=1\linewidth]{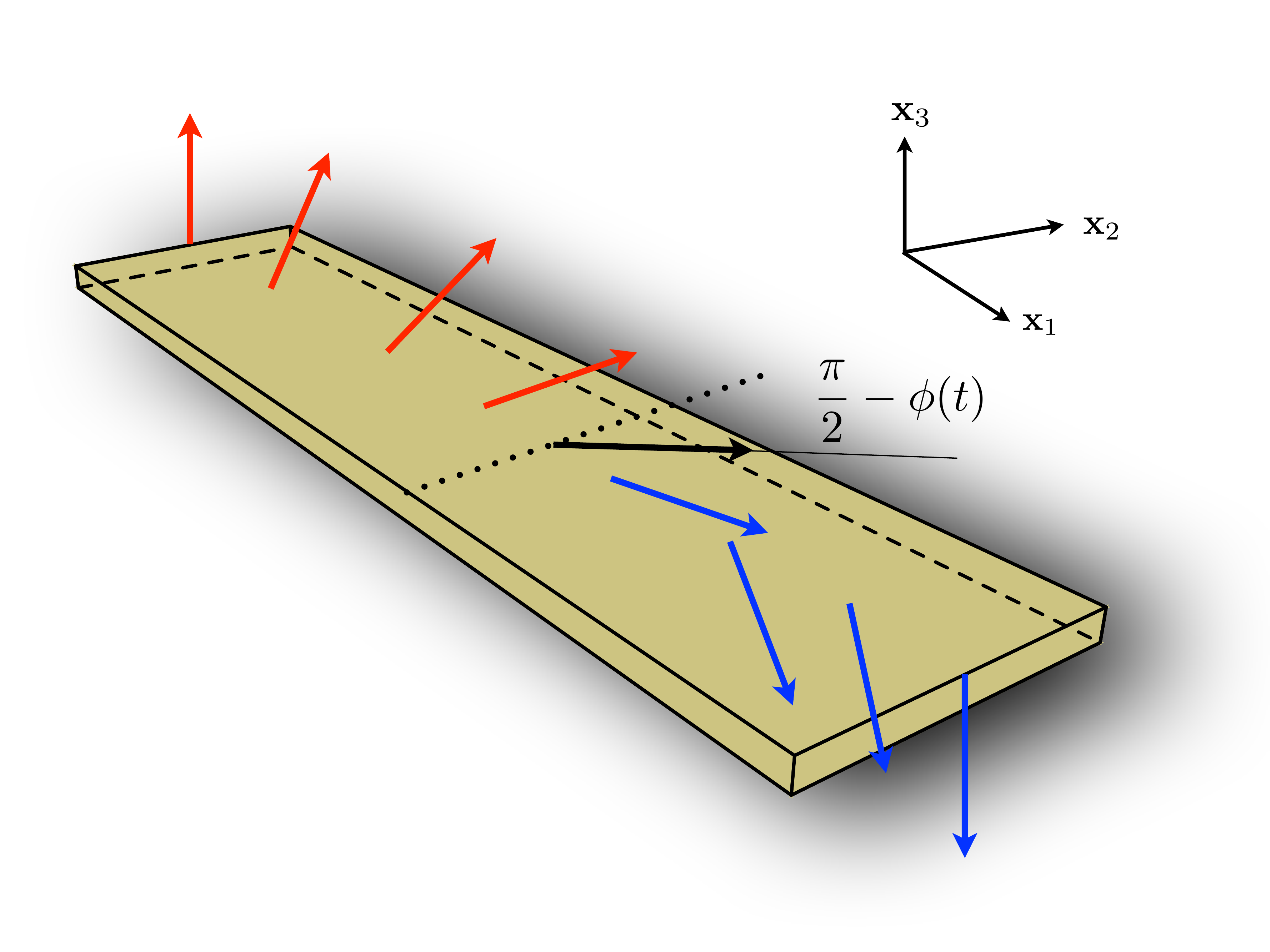}%[angle=0,width=0.8\linewidth]
\end{center}
\caption{(Color online) Artistic image of the DW considered in the text. The dotted line defines $X(t)$ and the continuous line defines $\phi(t)$. Red (light) color means a positive out-of-plane magnetization while blue (dark) color means a negative out-of-plane magnetization. The black arrow indicates a magnetization pointing in the $\mathbf{x}_{1}-\mathbf{x}_{2}$ plane.}
\label{figart}
\end{figure}

In this context, three dimensional Topological Insulators (TI) provide a new spectrum of possibilities regarding DW motion. An interesting aspect of the surface states of three dimensional TI is that external magnetic agents can modify their spectrum\cite{KGV13} and electronic transport properties\cite{WKF13}, and conversely, the dynamics of magnets coupled to these surface states can be severely modified\cite{FG10,NN10,YZN10,TL12}. Specifically, the question of how the presence of these surface states influences the dynamics of a DW has been recently addressed in the literature\cite{NN10,TL12}. 
We will go further by bringing into play an out of plane magnetic field, and show how magnetic field generated DW velocity can be significantly increased by holding off the appearance of Walker breakdown via an external electric field, and how the macroscopic DW chirality can be controlled by the presence of the TI surface state. In the way, previous unnoticed terms in the DW effective action will be obtained.

The paper is organized as follows: In section \ref{sec:model} we present the model for both the magnetic thin film and the TI surface states. The DW solution is presented together with the description of the coupling between the DW and the TI surface states. In section \ref{sec:effaction} we obtain the effective action and equations of motion of the DW in terms of collective coordinates by integrating out the fermionic degrees of freedom. In section \ref{sec:EDWestab} we describe how the presence of both magnetic and electric fields can stabilize the DW motion delaying the appearance of the Walker breakdown. In section \ref{sec:Currentcoherence} we describe how the DW chirality can be tuned at will by manipulating the electric and magnetic fields. At the end, in section \ref{sec:conclussions} we present a brief summary of the results obtained in this work.
\section{The model}
\label{sec:model}
To be specific we will consider an insulating ferromagnetic thin film, hosting a DW with an out of plane magnetization (see fig.(\ref{figart})), deposited on top of the surface of a three dimensional TI. The continuum Hamiltonian for the surface states of the TI, coupled by exchange interaction to the magnetic thin film, takes the standard form ($\hbar=1$ will be used):

$$
H_{TI}^\pm=\int d^{2}x\Big( v_{F} \psi^{+}\hat{\mathbf{x}}_3\cdot(i\nabla\times\bm\sigma)\psi \pm
$$
\begin{equation}
\pm\Delta \mathbf{m}(\mathbf{x},t)\cdot\psi^{+}\bm\sigma\psi\Big), \label{TIham}
\end{equation}
where $\Delta$ is the exchange coupling between the magnet and the surface states (it is definite positive) and the total magnetization $\mathbf{M}(\mathbf{x},t)$ relates with $\mathbf{m}(\mathbf{x},t)$ as $\mathbf{M}(\mathbf{x},t)=\gamma_{MI}/a^3\mathbf{m}(\mathbf{x},t)$. Here $\gamma_{MI}=\mu_{B}g_{MI}$ is the gyromagnetic ratio of the magnetic insulator ($\mu_B$ is the Bohr magneton and $g_{MI}$ is the Land\'e factor of the magnetic insulator). It is useful to write $\mathbf{m}(\mathbf{x},t)$ in spherical coordinates:
\be
\mathbf{m}(\mathbf{x},t)=S(\sin\theta\cos\phi,\sin\theta\sin\phi,\cos\theta)
\ee
The signs $+$ and $-$ in $H_{TI}^\pm$ correspond to antiferromagnetic and ferromagnetic exchange coupling, respectively. Regarding the external electromagnetic fields, we will analyze the effect of a magnetic field $B$ pointing along the $\hat{\mathbf{x}}_3$ direction, and an electric field $E$ pointing along the direction defined by the DW. These fields will couple to the TI surface states not only through the electromagnetic vector field $A_{\mu}$ but also through a Zeeman coupling $H_{z}=-\int d^{2}x\gamma_{TI}B \psi^{+}\sigma_{z}\psi$, being $\gamma_{TI}=\mu_{B}g_{TI}$ the gyromagnetic ratio for the topological insulator. For the magnetic layer, we will assume that its dynamics are conveniently described by an anisotropic Heisenberg model with an easy axis along $\hat{\mathbf{x}}_3$ and a hard axis on the $\hat{\mathbf{x}}_1-\hat{\mathbf{x}}_2$ plane\cite{STK11,TKS08}:

$$
H_{MI}=\frac{1}{2}\int \frac{d^3x}{a^3}\Big( J (\nabla \mathbf{m})^{2}+K^\bot_{x_1}m^{2}_{1}+K^\bot_{x_2}m^{2}_{2}-
$$
\be
-Km^{2}_{3}-\gamma_{MI} Bm_{3}\Big),\label{MIham}
\ee
where $J$ stands for the stiffness constant, and $K$, $K^\bot_{x_1,x_2}$ are the easy and hard axis anisotropies respectively. The full Lagrangian includes the dynamical term generated by the Berry phase of the spins\cite{STK11,TKS08}:
\be
\mathcal{L}_{MI}=\int \frac{d^3x}{a^3}\dot{\phi}S(\cos\theta-1)-H_{MI},\label{MIlag}
\ee
It is well known that the equations of motion derived from (\ref{MIlag}) support static extended solutions in the form of DWs \cite{S72}. Also in this situation, the excitation spectrum of the magnetic layer consists on gaped spin waves together with two zero energy modes. The later are related to invariance under translations of the DW center and rotations of the azimuthal angle that describes the DW chirality. Actually, the excitation associated to the azimuthal angle is a zero mode only if the hard axis anisotropy is zero, however one can consider small anisotropies so that it effectively decouples from the other massive excitations and still describes correctly the low energy dynamics\citep{STK11,TKS08}. To study the DW dynamics at low energies the zero modes are promoted to dynamical variables with finite kinetic energy. When the magnetic field is switched on or other external elements like electric field or the TI surface states are present, these dynamical variables are no longer zero modes, and become gaped excitations. Still, similar to the case with the hard axis anisotropy, they correctly describe the low energy dynamics if the energy remains smaller than that of the gaped excitation spectrum. 

The time evolution of the DW can be described in terms of the two variables $X(t)$ and  $\phi(t)$, representing translations and rotations of the azimuthal angle. The $x_3$ component of the magnetization takes the following form:

\begin{eqnarray}
m^{DW}_{3}&=&-S \tanh \left(\frac{x_1+X(t)}{\delta}\right),\label{magnetizationz}
\end{eqnarray}
where $\delta=\sqrt{J/K}$ is the DW width and $S$ is the maximum value of the magnetization. The in-plane components of $\mathbf{m}$ will be:
\begin{equation}
\mathbf{m}_{\perp}=Ssech\left(\frac{x_1+X(t)}{\delta}\right)\left(\cos\phi(t),\sin\phi(t),0\right)
\end{equation}

In order to find the effective theory describing the time evolution of the coordinates $X(t)$ and $\phi(t)$ we integrate out the fermionic degrees of freedom in the presence of the background field $m_3^{DW}$. To do this we use the fermionic spectrum which has been extensively discussed in the literature\cite{CH85,C94}. It is important to note that, besides the third component of the magnetization acting as a mass for the TI surface states, we also have two other components for $\mathbf{m}^{DW}$. These components are coupled to the other two Pauli matrices in the fermionic action (\ref{TIham}), which of course do not commute with $\sigma_{3}$ and thus induce couplings between the fermionic eigenstates previously calculated. The technical details of this derivation are presented in sections \ref{app:boundeff}, \ref{app:scatteff}, \ref{app:Bfieldaction} of the Appendix, here we will comment on the most salient features of the calculation. First of all, the presence of an external magnetic field $B$ pointing along the perpendicular direction to the TI surface states completely changes the fermionic spectrum (and hence the effective action) depending on if $|B|$ is larger or smaller than $S\Delta/\gamma_{TI}$. The reason is mainly the appearance in the latter case of a fermionic chiral state where the mass $m\equiv \Delta m^{DW}_{3}+\gamma_{TI}B$ changes sign. In this regard we can define two regimes:

(i)\textit{Chiral regime}. It occurs when $|B|<S\Delta/\gamma_{TI}$, so the mass changes sign. The fermionic spectrum consists on a chiral massless state bound to (localized on) the wall, plus massive scattering (extended) states. There can appear massive bound states too. Specifically, the number of bound states (taking into account also the chiral state) is going to be the largest integer less than $\Delta \delta S/(2v_{F})+1$ (see Appendix \ref{app:boundeff}).

(ii)\textit{Non-chiral regime}. It occurs when $|B|>S\Delta/\gamma_{TI}$, so the mass does not change sign. As a consequence there is no chiral massless state. The spectrum consists on massive scattering states and, depending on the slope of $m^{DW}_{3}$, massive bound states.
%The effect of TI surface states in N\'{e}el DWs in the non-chiral regime has been studied previously in\cite{NN10}. 

The computation of the exact spectrum for a mass of the type $\tanh(x)$, and for any values of $\Delta, S,\delta$ and $v_F$ is a formidable task, and closed solutions are known only for special values of the parameters \cite{C94}. To obtain a solution for generic values, we will follow an approximate but much simpler route\cite{FL99}. We will approximate $m^{DW}_{3}$ to a straight line to obtain and integrate out the bound fermionic spectrum, while we will rely on an adiabatic approximation to integrate out the scattering states.

\section{Effective action for the DW dynamics}
\label{sec:effaction}
By integrating out the fermionic degrees of freedom (both massless and massive) up to one loop level and adding the part coming from the isolated MI, we obtain the effective action for the collective coordinates $X(t)$ and $\phi(t)$ (see sections \ref{app:boundeff}, \ref{app:scatteff}, \ref{app:Bfieldaction} of the Appendix for details on the computation): 
\be
\Gamma_{DW}^{\pm}=\Gamma_{MI}+\Gamma_{TI}^{\pm},
\label{eq. effective action}
\ee
The MI part is:
\begin{widetext}
\be
\Gamma_{MI}=S\mathcal{N}\int dt\int_{-\frac{L_2}{2}}^{\frac{ L_2}{2}} dx_2\Big\{\frac{\dot{X}}{\delta}\phi-\frac{K_{x_1}^{\bot}S}{2}\cos^2\phi-\frac{K_{x_2}^{\bot}S}{2}\sin^2\phi+\gamma_{MI}B\frac{X}{\delta}\Big\},
\ee
\end{widetext}
where $\mathcal{N}$ is the number of spins in the wall divided by $L_2$. For the TI part, the calculation is different depending whether we are in the chiral or non-chiral regime. We shall present the results separately for both situations.

\subsection{Effective action in the chiral regime}

In this regime $|B|<S\Delta/\gamma_{TI}$. Setting $S=1$, an exchange coupling of $\Delta=0.1\,eV$, a Land\'e factor for the topological insulator of $g_{TI}=100$, and restoring $\hbar$ we find that this regime is defined for values of $B$ up to $|B|<17T$ (the reason for such quite a big, although still realistic, exchange coupling will be transparent in section \ref{sec:EDWestab}). For so large magnetic fields we should consider both the orbital coupling that originates the formation of Landau levels (LL), and the Zeeman coupling, which generates an splitting of these levels. The stronger the magnetic field, the more important the Zeeman mass term is in comparison with the energy of the first non zero LL, which scales with the magnetic field as $B^{1/2}$.  In typical 3D TIs, however, the Zeeman splitting remains much smaller than the energy separation of the LL even for quite big fields. As an example, the Zeeman splitting is negligible for fields up to at least $B=11T$ in Bi$_2$Se$_3$\cite{CSZ10}.

At low fields, far from the formation of LL, we can treat the orbital coupling perturbatively. In this situation the Zeeman coupling will contribute to a parity breaking mass term of the form $m=\Delta m^{DW}_{3}+\gamma_{TI}B$, which will give rise to a Quantum Anomalous Hall Effect (QAHE) with a topological Chern-Simons term\cite{R84,QWZ06}.

For fields larger than $B\approx 2T$ the LL formation becomes relevant\cite{CSZ10}. At these fields, in order to obtain the effective action at low energies it is enough to consider the ultra quantum limit, where only the lowest LL is populated and  inter Landau level transitions are neglected. In this situation the mass term will be equal to that of the QAHE just described $m=\Delta m^{DW}_{3}+\gamma_{TI}B$ but now we are in the scenario of a normal Quantum Hall Effect (QHE). As long as we only  consider the zero LL, the filling factor is just 1, and a Chern-Simons term analogous to that of the QAHE is obtained\cite{H84}. 

We conclude that the same effective field theory describes both situations of small and large magnetic fields. We have to stress that this considerations have to do only with the massive scattering fermions, which live in (2+1) dimensions. The chiral state is not modified when changing the magnetic field due to its chiral nature. The massive bound states, however, change when varying $B$ but they do not generate a topological response. So for the bound states a perturbative treatment of the orbital contribution is justified, as in the case of a weak magnetic field.

All considered, after computing the spectrum for the corresponding mass term and integrating out the fermions we arrive to (in real time):
\begin{widetext}
\be
\Gamma_{TI}^{\pm}=S\int dt\int_{-\frac{ L_2}{2}}^{\frac{ L_2}{2}} dx_2\Big\{-\frac{\Delta}{v_F} J^{EM}_{2,\pm}\cos\phi-\frac{\mathcal{N}K_0^{\bot}S}{2}\cos^2\phi-\frac{\mathcal{N}K_m^{\bot}S}{2}\sin^2\phi\pm\gamma_{eff}\delta BE\sin\phi\Big\}.
\label{eq. effective action TI1}
\ee
\end{widetext}
It is worth pointing out that terms 2,3 and 4 of the r.h.s. of eq. (\ref{eq. effective action TI1}) are new, in the sense that they were not previously reported in the literature. The one dimensional (integrated over $x_1$) electromagnetic current density flowing along the DW, $J^{EM}_{2,\pm}$, is (see appendix \ref{app:emcurrent}):
\be
J^{EM}_{2,\pm}=\frac{v_F}{2\pi}\frac{1}{\partial_0\pm v_F\partial_2}E.
\ee
Notice that $J^{EM}_{2,\pm}$ is written in terms of the integral operator $1/(\partial_0\pm v_F\partial_2)$. An specific form of the electromagnetic current has to be obtained by fixing the boundary conditions, in order to study the physics in the chiral regime (Appendix \ref{app:emcurrent}). $K_0^{\bot}=\Delta^2/(2\pi \mathcal{N} v_F)$ and $\gamma_{eff}=\gamma_{TI}/(2\pi v_FS)$, and we do not have an exact value for $K_m^{\bot}$, although we know it is of the order of $K_0^{\bot}$ (see Appendix \ref{app:boundeff}). $K_0^{\bot}$ and $K_m^{\bot}$ renormalize the MI hard axis anisotropies $K_{x_1}^{\bot}$ and $K_{x_2}^{\bot}$ respectively, however their value is going to be considerably smaller than their MI counterparts. Hence the uncertainty in the value of $K_m^{\bot}$ will have no impact on the final results. It is worth to notice that $K_0^{\bot}$ is generated by the chiral mode, while $K_m^{\bot}$ is generated by the coupling between the chiral and massive modes.

We see that the fermionic fluctuations contribute with a torque term proportional to $\cos\phi J^{EM}_{2,\pm}$ generated by the coupling with the chiral mode. This torque can be understood in terms of the non conservation of the one dimensional chiral current flowing along the DW. According to the Callan-Harvey mechanism\cite{CH85} the non conservation of the chiral current is compensated by a charge inflow from the two dimensional bulk. The spin of the chiral state, being perpendicular to the current motion is forced to lie on the $\mathbf{x}_{1}-\mathbf{x}_{2}$ plane. However, the current flowing from the bulk is spin polarized, and its polarization in the third direction is proportional to the sign of the induced mass. So the inflow current must change its spin polarization when reaching the chiral current, changing its total angular momentum. This excess of angular momentum is absorbed by the magnetic moments of the DW, generating the torque term.

We see also that when both magnetic and electric fields are switched on, the scattering fermions generate a term $\pm\gamma_{eff}\delta BE\sin\phi$ which stems from the Chern-Simons term. Finally and for completeness, we point out that the anisotropy term $K_m^{\bot}$ is renormalized as the number $\theta=\Delta \delta S/(2v_{F})$ increases, this is whenever an additional massive bound state appears. This renormalization is discontinuous, as for $0<\theta\leq 1$ the number of massive bound states $N$ is zero, for  $1<\theta\leq 2$ $N=1$, and so on (see Appendix \ref{app:boundeff} for details). For given TI and MI materials, massive bound states appear as the DW width $\delta$ increases. But as we already said, the exact value of $K_m^{\bot}$ is going to be irrelevant for us, as it is going to be small compared to $K_{x_1}^{\bot}$ and $K_{x_2}^{\bot}$.

The equations of motion are readily obtained from (\ref{eq. effective action TI1}) in this regime. After fixing the electromagnetic current configuration (see Appendix \ref{app:emcurrent}) and incorporating Gilbert damping\cite{STK11} we have:
\begin{widetext}
\be
\dot{\phi}=\frac{\gamma_{MI}B}{1+\alpha^2}+\frac{\alpha}{\mathcal{N}(1+\alpha^2)}\left(\frac{\Delta I_\pm}{v_F}\sin\phi-\frac{\mathcal{N}K^{\bot}S}{2}\sin{2\phi}\pm\gamma_{eff}\delta BE\cos\phi\right),
\label{eq. phi1}
\ee
\be
\dot{X}=\frac{\delta\gamma_{MI}}{\alpha}B-\frac{\delta}{\alpha}\dot{\phi},
\label{eq. X1}
\ee
where $I_\pm$ is the static (and spatially averaged) electromagnetic current configuration flowing along the DW, $\alpha$ is the Gilbert damping constant and $K^{\bot}=K_{x_2}^{\bot}+K_m^{\bot}-K_{x_1}^{\bot}-K_0^{\bot}$.
\end{widetext}

\subsection{Effective action in the non-chiral regime}

In this regime $|B|>S\Delta/\gamma_{TI}$, which is reached for values of  $|B|>17T$ for typical values of $\Delta\sim 0.1$ $eV$. However, for materials with exchange couplings two orders of magnitude smaller ($\Delta\approx 1meV$) this regime would become experimentally important ($|B|>0.17T$). Hence we will present the computation of the effective action in the non-chiral regime also for completeness.

The same analysis done in the previous section can be done here. The only difference is now the absence of the chiral state. After doing the corresponding computation (see Appendix \ref{app:Bfieldaction}) we get:
\be
\Gamma_{TI}^{\pm}=\pm\int dt\int_{-\frac{ L_2}{2}}^{\frac{ L_2}{2}} dx_2\,\frac{\Delta S\delta\,Sign(B)}{4v_F}E\sin\phi
\label{eq. effective action TI2}
\ee
We see that all the terms where the chiral fermions were involved disappear, and the only term remaining is the topological Chern-Simons term, but with a different value than in the chiral regime. This is so because in the chiral regime the mass changes sign at some point on the $\hat{\mathbf{x}}_1$ axis, while in the non-chiral regime the mass sign depends only on the sign of $B$ (see Appendix \ref{app:Bfieldaction}). 

The equations of motion now read:
\begin{widetext}
\be
\dot{\phi}=\frac{\gamma_{MI}B}{1+\alpha^2}+\frac{\alpha}{\mathcal{N}(1+\alpha^2)}\left(-\frac{\mathcal{N}K^{\bot}S}{2}\sin{2\phi}\pm\frac{\Delta\delta\,Sign(B)}{4v_F}E\cos\phi\right),
\label{eq. phi2}
\ee
\be
\dot{X}=\frac{\delta\gamma_{MI}}{\alpha}B-\frac{\delta}{\alpha}\dot{\phi},
\label{eq. X2}
\ee
with $K^{\bot}=K_{x_2}^{\bot}-K_{x_1}^{\bot}$.
\end{widetext}

\section{Electric field mediated DW chirality stabilization}
\label{sec:EDWestab}

\begin{figure*}
(a)
\begin{minipage}{.46\linewidth}
\includegraphics[scale=0.25]{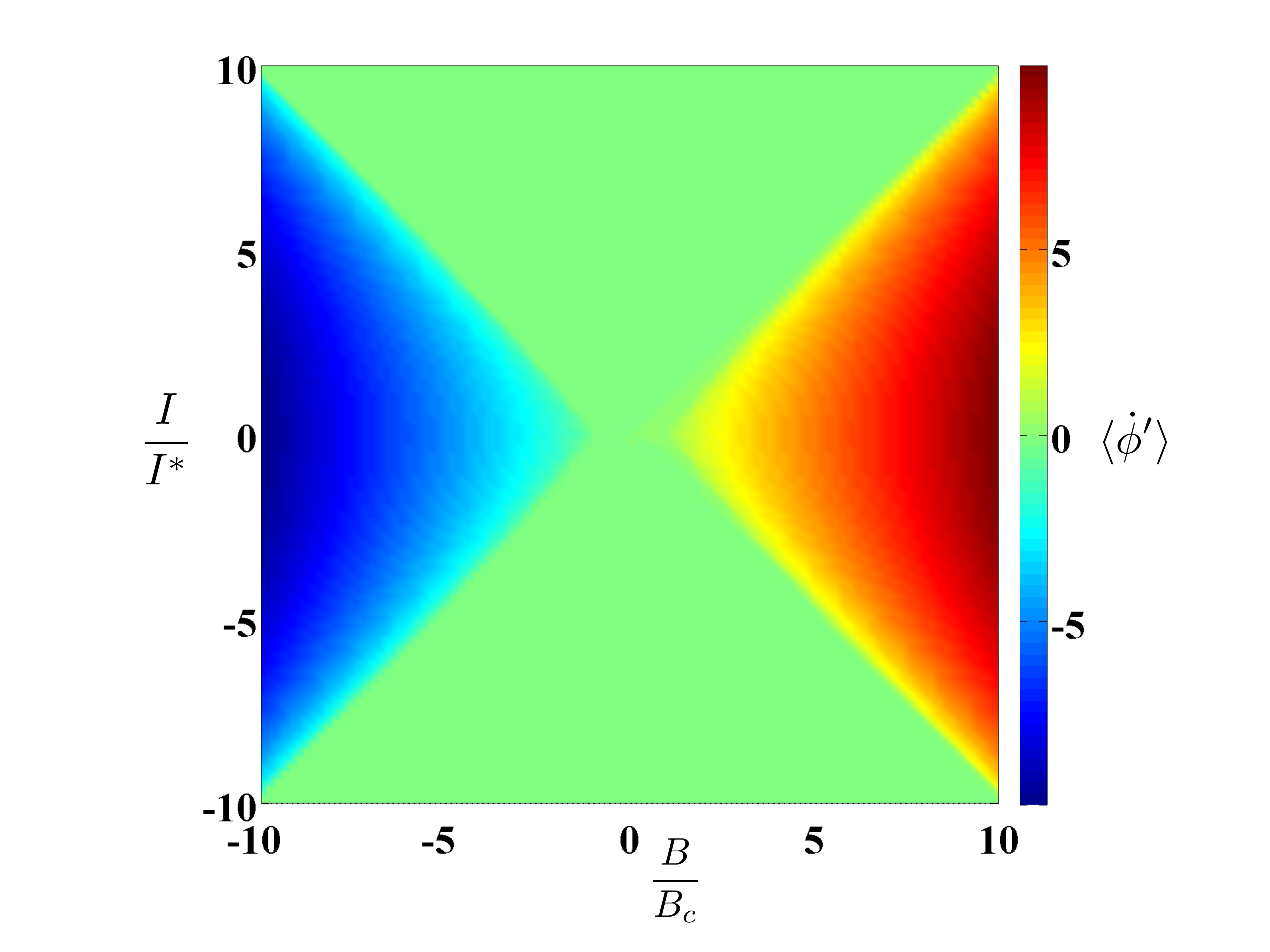}
\end{minipage}
(b)
\begin{minipage}{0.42\linewidth}
\includegraphics[scale=0.26]{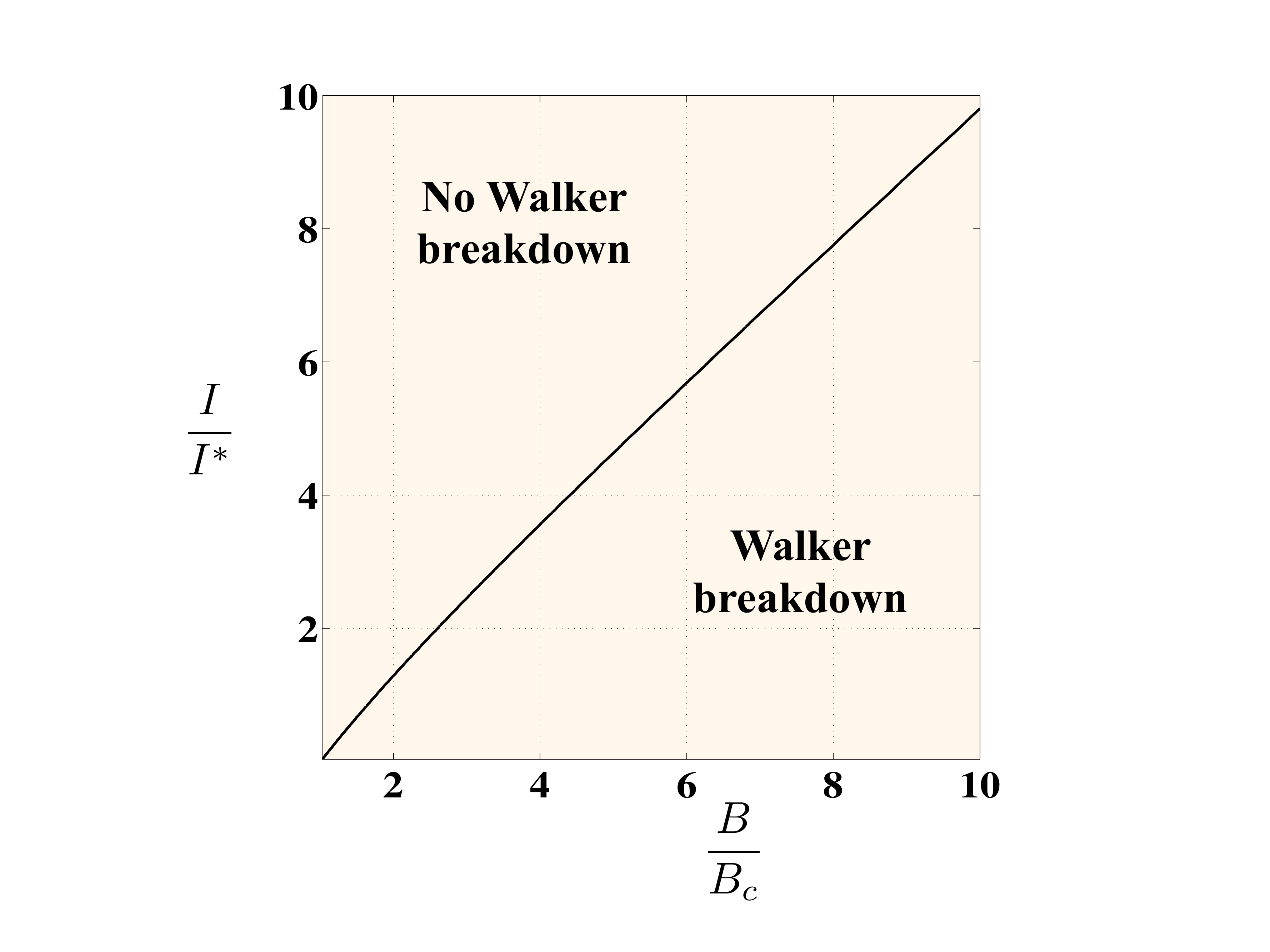}
\end{minipage}
\caption{\label{EvsB}(Color online) (a)$\langle\dot{\phi'}\rangle$ as a function of $I/I^*$ and $B/B_c$, valid for both ferro and antiferro exchange couplings. In the green flat region where $\langle\dot{\phi'}\rangle=0$ WB is absent. In the two triangular regions with $\langle\dot{\phi'}\rangle\neq0$ WB occurs. (b) WB frontier as a function of $I/I^*$ and $B/B_c$. Again valid for both ferro and antiferro exchange couplings.}
\end{figure*}

If we turn off the electric field, both in the chiral and non-chiral regime the equations of motion are equivalent to those of a DW in a MI in the presence of an easy-axis magnetic field, with well known solutions\cite{STK11}. For magnetic fields smaller than a critical field $B_c=\alpha K^\bot S/(2\hbar\gamma_{MI})$, the time-averaged terminal velocity of the Wall is $\langle\dot{X}\rangle=\mu B$, where $\mu$ is the mobility and its value is $\mu=\delta\gamma_{MI}/\alpha$. When $B$ reaches $B_c$ Walker breakdown (WB) occurs and $\phi$ starts to precess decreasing the terminal velocity as $B$ increases. If $B$ continues growing there is a point where the terminal velocity starts to increase linearly again but with considerably lower mobility\cite{STK11}. Thus being able to stabilize the chirality $\phi$ avoiding this WB regime is essential to maintain the high initial mobility $\mu=\delta\gamma_{MI}/\alpha$ and hence to reach high velocities.

As we pointed out in section \ref{sec:effaction}, for values of $\Delta=0.1eV$ and $g_{TI}=100$, the chiral regime extends up to fields of $B=17T$ so we will always be in this regime. After switching on the electric field, the picture is modified with respect to that of an isolated MI as two new terms appear in the equations of motion (terms 2 and 4 of the r.h.s of eq. (\ref{eq. phi1})). These new terms will act as chirality stabilizers, delaying the appearance of WB. The equations of motion for the chiral regime can be written in a more manageable form:
\be
\dot{\phi'}=\frac{B}{B_c}-\sin{2\phi}\mp\frac{I}{I^*}\sin\phi\pm r \frac{I}{I^*}\frac{B}{B_c}\cos\phi,
\label{eq. phi11}
\ee
\be
\dot{X}=\mu B-\frac{\delta}{\alpha}\dot{\phi},
\label{eq. X11}
\ee
with:
\be
I=-\frac{e^2}{2h}V;\quad I^*=\frac{\mathcal{N}v_FeK^{\bot}S}{2\Delta}
\ee
\be
\phi'=\frac{2\hbar(1+\alpha^2)}{\alpha K^{\bot}S}\phi
\ee
\be
r=\frac{2\hbar\gamma_{TI}\delta B_c}{L_2\Delta S}.
\ee
$V$ is the voltage between both sides of the magnetic strip\citep{TL12} (see Appendix \ref{app:emcurrent}). Notice that we reintroduced $\hbar$ and $e$.

To make a quantitative analysis, we need to give values to the different parameters appearing in the model.  For a permalloy MI strip, we can set a DW width of $\delta=10nm$, a Gilbert damping constant of $\alpha=0.01$, a Landee factor of $g_{MI}=2$, a strip thickness of 1nm, a lattice constant $a=0.35 nm$ and we do $S=1$. The density of spins of the wall in the $\hat{\mathbf{x}}_2$ direction is $\mathcal{N}=2.3\times 10^{11}m^{-1}$. For the hard axis anisotropy energy we set\cite{STK11} $K^{\bot}=1\kappa_B1K$. On the other hand, for the TI we set a Fermi velocity of $v_F=5\times 10^5$. The TI Land\'e factor and the exchange coupling energy have been already introduced.

For these values of the parameters we have $I^*=8\times 10^{-6}A$, $B_c=3.8\times10^{-3}T$, $\mu=1.9\times10^{5}T^{-1}m/s$ and $r=4.4\times10^{-12}/L_2$. Such a small value for $r$ means that for strip widths down to $L_2\sim1\mu m$, the last term of the right hand side of equation (\ref{eq. phi11}) can be neglected for all possible values of the field up to which the chiral regime extends, and the chirality stabilization is mediated  by the chiral current (third term) rather than by the topological response generated by the extended (2+1) dimensional states (last term). 

We plot in fig. (\ref{EvsB}a) the averaged terminal precession velocity $\langle\dot{\phi'}\rangle$, obtained by numerically solving eq. (\ref{eq. phi11}), as a function of $B/B_c$ and $I/I^*$. The green (light) region indicates zero average precession velocity corresponding to the non WB regime. In fig. (\ref{EvsB}b) we plot the border between the two regimes for positive $I$ and $B$ for which the magnetic field acquires its critical value $B=B_c$. It is apparent how $B_c$ increases with $I$ with a close to linear behavior, so we face an scenario where the corresponding high mobility $\mu$ of the non WB region is extended to higher fields. This way high velocities can be achieved with relatively low magnetic fields. We stress that the behavior is the same with both ferro and antiferromagnetic exchange couplings.

Let us take a look to the value of the current $I^*$. If we set a magnetic strip width of $L_2=100\mu m$, this current would give rise to a current density of $8\times10^{-2}A/m$. This is approximately an order of magnitude smaller than the value at which the breakdown of the dissipationless current occurs ($\sim1A/m$)\cite{N99,VM09}. Looking at fig. (\ref{EvsB}), for a magnetic strip $100\mu m$ wide it would be challenging to stay in the non WB region for fields higher than $B=10B_c$, as one would need currents of at least $I\sim10I^*$, which equals to $0.8A/m$ or higher. The wider the strip, the lower the current density corresponding to the current $I^*$, so the situation improves for wider strips. For a field $B=10B_c$ and a current density of $0.8A/m$, a velocity of $\dot{X}=7.2\times10^{3}m/s$ is achieved for a strip $100\mu m$ wide, an order of magnitude higher than that at $B_c$, which is a promising result.

\section{Current induced DW chirality reversal and macroscopic quantum coherence of chirality}
\label{sec:Currentcoherence}

\begin{figure*}
%\begin{center}
\includegraphics[angle=0,width=0.8\linewidth]{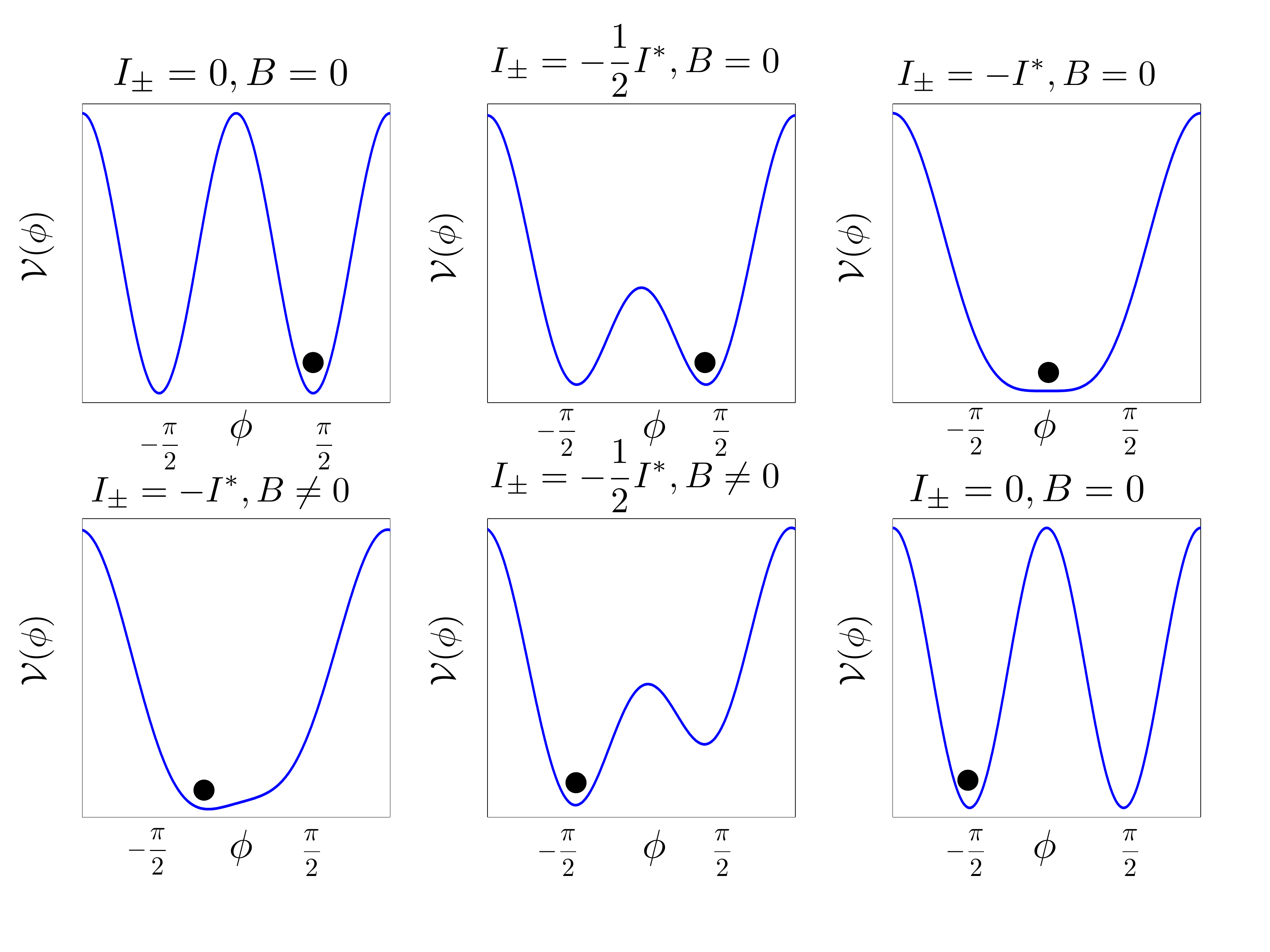}%[angle=0,width=0.8\linewidth]
%\end{center}
\caption{Schematic sequence of chirality reversal (from left to right and top to bottom). The black point represents the actual DW chirality.}
\label{fig4}
\end{figure*}

In this section we will consider again the chiral sector, initially at zero magnetic field. Solving the equation for $\phi$ we find that as $t$ increases $\dot{\phi}$ tends to zero and the angle $\phi$ stabilizes. There are two cases to consider, $sign(K^{\bot})=\pm1$ which give qualitatively different potential energies. The case of interest for us is $sign(K^{\bot})=-1$ occurring when $K_{x_2}^{\bot}=0$ and assuming that $K_{x_1}^{\bot}$ is much larger than the anisotropies generated by the fermionic fluctuations. In this case, the potential energy density is:
\be
\mathcal{V}(\phi)=\frac{S\mathcal{N}K^\bot_{x_1}}{2}\left(\cos^2{\phi}+\frac{I_\pm}{I^*}\cos{\phi}\right),
\label{eq. potential}
\ee
We see that the vacuum is degenerate. If $|I_\pm|\geq 2I^*$ the vacuum is given by $\phi=2n\pi$ for $I_\pm<0$ and $\phi=(2n+1)\pi$ for $I_\pm>0$. On the other hand, if $|I_\pm|<2I^*$ an splitting occurs so there are two minima $\phi_\pm$ for each integer $n$ at $\phi_\pm=2n\pi\pm\arccos{(-I_\pm/(2I^*))}$ for $I_\pm<0$ and $\phi_\pm=(2n+1)\pi\pm\arccos{(-I_\pm/(2I^*))}$ for $I_\pm>0$. These two minima are separated by a potential barrier whose height increases as $|I_\pm|$ decreases towards zero. In fig. (\ref{fig4}), the potential is plotted for different values of $I_\pm$ (see the cases with $B=0$ only).

There is a way to flip the chirality of the Wall with the help of a weak magnetic field. It is schematically shown in fig. (\ref{fig4}). The idea is to start with a wall configuration in the absence of current, sitting at $\phi=\pi/2$ which is a minimum of the potential for $I_\pm=0$. Then we switch on a current and slowly increase it up to $-2I^*$, flipping the angle $\phi$ in the process down to zero. At this point, we introduce a small magnetic field in the $\hat{\mathbf{x}}_3$ direction, and start to decrease the current towards zero. An extra term $\pm\gamma_{eff}\delta BE\sin\phi$ is generated (see eq. (\ref{eq. effective action TI1})), such that the reflection symmetry protecting the degeneracy of the two minima $\phi_+$ and $\phi_-$ is broken. Now by choosing the correct sign for $B$, $\phi$ will fall to $\phi_-$, and one can turn off the magnetic field again. Continuing to decrease the current, we finally end up at $\phi=-\pi/2$ when $I_\pm=0$. This way we have flipped the chirality of the DW in a totally controlled manner.

Assuming that this flip is performed adiabatically (with a sufficiently slowly varying current), in the process the wall performs a displacement of $\Delta X=\pi\delta/\alpha$. This displacement would be an experimental signature of the chirality reversal\cite{ITH11}.

Let us now introduce a strong enough pinning potential so we can integrate out the coordinate $X$ and treat $\phi$ as a particle moving in the potential given by eq. (\ref{eq. potential}). Quantum tunnelling of chirality between different vacua is possible. The frequency of the quantum coherent oscillation between two first neighbor vacua has been obtained before, for a DW in a ferromagnet with finite hard axis anisotropy\cite{TST96}. Their configuration is analogous to ours if we do $I_\pm=0$, the two vacua corresponding to $\phi=\pi/2$ and $\phi=-\pi/2$. In ref.\cite{TST96} it is found that strong pinning and weak hard axis anisotropy favour the quantum coherence between the two chiralities. In our configuration, the potential barrier that separates the two $\phi_\pm$ vacua decreases as the absolute value of the current is increased towards $2I^*$, and disappears when it reaches $2I^*$. As the probability of quantum tunnelling decreases exponentially with the height of the barrier, the quantum coherence between $\phi_+$ and $\phi_-$ can be significantly enhanced by the application of a current. This way one can significantly rise the frequency for the quantum coherent oscillation between positive and negative chiralities, making this phenomenon easier to be observed.

\section{Conclussions} 
\label{sec:conclussions}
In the present work we have obtained the equations of motion for a DW coupled both to the surface states of a TI and to external electromagnetic fields. The external electric field acts on the DW via the chiral state through a new type of spin-torque mechanism. By controlling both the electric and magnetic fields the appearance of the Walker breakdown can be hold off significantly increasing the terminal velocity of the DW. Also through the appropriate tuning of the electromagnetic fields one can reverse the DW chirality in a controllable manner. This control can be of use to design future logic gates.   

\section{Acknowledgments}
The authors gratefully acknowledge M. A. H. Vozmediano for valuable comments and suggestions. This research is partially supported by CSIC JAE-doc fellowship program and the Spanish MECD Grants No. FIS2011-23713 and No. PIB2010BZ-00512.
\appendix
\section{Contribution of the fermionic bound states to the effective action for B=0}
\label{app:boundeff}
Here we show how the computation of the contribution of the fermionic bound states is done, in the absence of magnetic field and for ferromagnetic coupling (it is straight forward to change the calculation for antiferromagnetic coupling). Next  we will do the computation for the scattering states, which will give a (2+1) dimensional topological Chern-Simons term as we will see. The remaining part would be the contribution of the terms arising from the coupling between the scattering and bound states, but it's computation is much more involved and we will skip it. To do so we acknowledge that the scattering states renormalize all terms generated by the massive bound states (via scattering-scattering and scattering-bound states couplings), or in other words the massive bound states can not generate terms additional to those generated by the scattering states. However, this is not true the other way around, this is, the massive bound states do not renormalize all terms generated by the scattering states. There is one term that is specific from the (2+1) dimensional extended states, which is the topological Chern-Simons term.

Then as we said, the coupling between the scattering and bound states is going to renormalize the terms generated by the bound states alone. By taking into account only the bound state contribution plus the (2+1) Chern-Simons term, we are going to obtain all the possible terms generated by the massive states, but are going to miss the renormalization to the non topological terms coming from the scattering-bound states coupling. However, this will not be important and will not have an appreciable impact in the final results, as we point out in section \ref{sec:effaction}.

As the bound states are localized around $x=-X$, let us expand $\mathbf{m}^{DW}(x_1+X)$ around this point up to the linear term\citep{FL99}. For the electromagnetic field we choose the gauge $A_{1}=0$ and $A_{i}=A_{i}(t,x_2)$, so we explore the effect of an electric field $E$ pointing along the $\hat{\mathbf{x}}_2$ direction and zero magnetic field $B$. Under these assumptions, and from the continuum Hamiltonian for the surface states of the TI (see main text), we can write the fermionic action as (all computations in the Appendix will be done in imaginary time, except those in section \ref{app:emcurrent}):
\begin{subequations}
\begin{eqnarray}
\mathcal{S}_{TI}^{bs}=\int dtd^2x\overline{\Psi}\Big(D_1(x_1)+D_2(t,x_2)\Big)\Psi, \label{TIaction}
\end{eqnarray}
\begin{eqnarray}
D_1(x_1)=-v_F\sigma_1\partial_1-\Delta S\frac{x_1+X}{\delta},\label{appeq1}
\end{eqnarray}
\begin{eqnarray}
\nonumber D_2(t,x_2)&=&\sigma_3(\partial_0-iA_0)-v_F\sigma_2(\partial_2-iA'_2)+\\
&+& iv_F\sigma_1A'_1, \label{appeq2}
\end{eqnarray}
\end{subequations}
with $A'_1=-\Delta S\sin\phi/v_F$, $A'_2=A_2+\Delta S\cos\phi/v_F$ and $\overline{\Psi}=\Psi^\dagger\sigma_3$. Notice that we did $A_\mu\rightarrow A_\mu/e$.

We will exactly compute the spectrum of the operator (\ref{appeq1}) and use it as a basis for the fermion fields. To this end we can express $D_{1}$ in terms of creation and annihilation operators of the harmonic oscillator and \emph{chiral} projectors:

\begin{equation}
D_{1}=\sqrt{\frac{2\Delta S v_F}{\delta}}(a P_{R}+a^{\dagger}P_{L}),\label{appeq3}
\end{equation}
where
\begin{subequations} 
\label{appeq4}
\begin{eqnarray}
a=-\sqrt{\frac{\delta}{2\Delta S v_F}}(v_F\partial_{1}+\frac{\Delta S}{\delta}(x_1+X)),
\end{eqnarray}
\begin{eqnarray}
a^{\dagger}=\sqrt{\frac{\delta}{2\Delta S v_F}}(v_F\partial_{1}-\frac{\Delta S}{\delta}(x_1+X)),
\end{eqnarray}
\begin{eqnarray}
P_{R}=\frac{1}{2}(1+\sigma_{1}),\quad P_{L}=\frac{1}{2}(1-\sigma_{1}).
\end{eqnarray}
\end{subequations}
$a$ and $a^\dagger$ fulfill the commutation relation $[a,a^\dagger]=1$.

$D_{1}$ is not an hermitian operator, so we diagonalize $D^{\dagger}_{1}D_{1}$ instead and obtain:
\begin{subequations}
\label{appeq5}
\begin{eqnarray}
\Psi(t,x_1,x_2) &=&\sum^{\infty}_{n=0}\lbrace\rho_{n}(x_1+X)\Psi^{n}_{R}(t,x_2)+\\\nonumber
&+&\rho_{n-1}(x_1+X)\Psi^{n}_{L}(t,x_2)\rbrace ,
\end{eqnarray}
\begin{eqnarray}
\overline{\Psi}(t,x_1,x_2)&=&\sum^{\infty}_{n=0}\lbrace\overline{\Psi}^{n}_{R}(t,x_2)\rho^{*}_{n}(x_1+X)+\\\nonumber
&+&\overline{\Psi}^{n}_{L}(t,x_2)\rho^{*}_{n-1}(x_1+X)\rbrace ,
\end{eqnarray}
\end{subequations}
with $\Psi_{R,L}^{(n)}=P_{R,L}\Psi^{(n)}$ and $\overline{\Psi}_{R,L}^{(n)}=\Psi^{(n)}P_{L,R}$. The functions $\rho_{n}$ are defined through ($\int dx\rho^{*}_{n'}\rho_{n}=\delta_{n'n}$):
\begin{subequations}
\begin{eqnarray}
a^{\dagger}a\rho_{n}=n\rho_{n}&,&aa^{\dagger}\rho_{n}=(n+1)\rho_{n},\\\nonumber
\end{eqnarray}
\begin{eqnarray}
a^{\dagger}\rho_{n}=\sqrt{n+1}\rho_{n+1}&,& a\rho_{n}=\sqrt{n}\rho_{n-1}.
\end{eqnarray}
\end{subequations}
From the standard knowledge of the theory of the harmonic oscillator, we know that there is always a zero mode $D^{\dagger}_{1}D_{1}\Psi_{0}=0$ with $\Psi_{0}=\rho_{0}\psi^{0}_{R}(t,x_2)$ being chiral. 

So far we have calculated the eigenstates linearizing $\tanh(\frac{x_1+X}{\delta})$, and we have an infinite discrete Hilbert spectrum. The linear approximation should only be valid in the region $|x_1+X|<\delta/2$. Beyond this point, the spectrum already calculated differs substantially to the real one, which is known to consist on a finite number of bound states and the continuum (scattering states), so we have to impose a cut off in $n$. There are two equivalent ways of doing it. One is to compute the spatial dispersion of each eigenstate in the coordinate $x_1$ and assume that it must be smaller than the DW width. The other possibility is to impose that the masses of the bound states have to be smaller than the asymptotic mass of the scattering states. Following the second route, we have that the highest value of $n$ is reached when:
\begin{equation}
\sqrt{D^{\dagger}_{1}D_{1}}\rho_{N}<\Delta S \rho_{N},
\end{equation}
so
\begin{equation}
N<\frac{\Delta S \delta}{2v_F}\equiv \theta.
\end{equation}
Taking into account the (massless) zero mode, the total number of bound states is $N+1$, being $N$ the largest integer smaller than $\theta$. To illustrate the effect of having more bound states than the chiral one, we will restrict ourselves to calculate the effective action for the two illustrative cases of $N=0$ (just the chiral zero mode) and $N=1$ (one massive bound state apart of the chiral zero mode).

At this point, it is convenient to explicitly decompose the bispinor $\Psi^{n}=\left(\alpha^{n},\chi^{n}\right)$ in its chiral parts:

\begin{subequations}
\label{appeq7}
\begin{eqnarray}
\Psi^{n}_{R}=\psi^{n}_{R}\frac{1}{\sqrt{2}}\begin{pmatrix}
1\\
1
\end{pmatrix}, \Psi^{n}_{R}=\psi^{n}_{R}\frac{1}{\sqrt{2}}\begin{pmatrix}
1\\
-1
\end{pmatrix}
\end{eqnarray}
\begin{eqnarray}
\psi^{n}_{R}=\frac{1}{\sqrt{2}}(\alpha^{n}+\chi^{n}),\; \psi^{n}_{L}=\frac{1}{\sqrt{2}}(\alpha^{n}-\chi^{n}).
\end{eqnarray}
 \end{subequations}
 
\begin{widetext}
Now, substituting in (\ref{TIaction}) the eigenmode expansion (\ref{appeq5}), using the decomposition (\ref{appeq7}) and integrating over the spatial coordinate $x_1$, we arrive at the following (1+1) action:

\begin{subequations}
\label{fermionicaction}
\begin{eqnarray}
\mathcal{S}^{bs}_{TI}=\mathcal{S}^{0}+\mathcal{S}^{1}+\mathcal{S}^{01},
\end{eqnarray}
\begin{eqnarray}
\mathcal{S}^{0}=\int dtdx_2\psi^{0+}_{R}(\hat{\partial}-i\hat{A})\psi^{0}_{R},
\end{eqnarray}
\begin{eqnarray}
\mathcal{S}^{1}=\int dtdx_2 \left\{\psi^{1+}_{R}(\hat{\partial}-i\hat{A})\psi^{1}_{R}+\psi^{1+}_{L}(\hat{\partial}^{*}+i\hat{A}^{*})\psi^{1}_{L}+\sqrt{\frac{2\Delta Sv_F}{\delta}}(\psi^{1+}_{R}\psi^{1}_{L}+\psi^{1+}_{L}\psi^{1}_{R})\right\},
\end{eqnarray}
\begin{eqnarray}
\mathcal{S}^{01}=\int dtdx_2\; \left\{ iv_F(\psi^{1+}_{L}A'_{1}\psi^{0}_{R}-\psi^{0+}_{R}A'_{1}\psi^{1}_{L})+\sqrt{\frac{\Delta S}{2v_F\delta}}(\psi^{1+}_{R}\dot{X}\psi^{0}_{R}-\psi^{0+}_{R}\dot{X}\psi^{1}_{R})\right\},
\end{eqnarray}
\end{subequations}
where we have defined the \emph{chiral} operators $\hat{\partial}=\partial_{0}+iv_F\partial_{2}$ and $\hat{A}=A_{0}+iv_FA'_{2}$, and $\dot{X}=\partial_0X$. It is important here to note that when $N=0$ ($0<\theta\leq 1$) only the chiral mode is present so $\mathcal{S}^{1}$ and $\mathcal{S}^{01}$ are absent. 

In order to get the effective action in terms of the variables $X(t)$ and $\phi(t)$, and the external electric field, we integrate out the fermions in the corresponding functional integral and find, to 1-loop order::

\begin{subequations}
\label{effaction}
\begin{eqnarray}
\Gamma^{bs}_{TI}=\Gamma^{bs}_{A'}+\Gamma^{bs}_{X}+\Gamma^{bs}_{||},
\end{eqnarray}
\begin{eqnarray}
\Gamma^{bs}_{A'}=-\frac{1}{2} Tr \left\lbrace \hat{\partial}^{-1}\hat{A}\hat{\partial}^{-1}\hat{A}+(\hat{\partial}g_{0}\hat{A}^{*}\hat{\partial}g_{0}\hat{A}^{*}+c.c.)+\frac{4\Delta Sv_F}{\delta}g_{0}\hat{A}g_{0}\hat{A}^{*}\right\rbrace,
\end{eqnarray}
\begin{eqnarray}
\Gamma^{bs}_{X}=-i\frac{\Delta Sv_F}{\delta} Tr \left\lbrace g_{0}A'_{1}\hat{\partial}^{-1}\dot{X}+\hat{\partial}^{-1}A'_{1}g_{0}\dot{X}\right\rbrace,
\end{eqnarray}
\begin{eqnarray}
\Gamma^{bs}_{||}=-v_F^{2}Tr\left\lbrace \frac{2\Delta Sv_F}{\delta}\hat{\partial}g_{0}A'_{1}\hat{\partial}^{-1}g_{0}A'_{1}-\hat{\partial}g_{0}A'_{1}\hat{\partial}^{*}g_{0}A'_{1}\right\rbrace,
\end{eqnarray}
\end{subequations}
where $Tr$ is the trace and $g_{0}=(\hat{\partial}\hat{\partial}^{*}-2\Delta Sv_F/\delta)^{-1}$. The evaluation of the integrals in (\ref{effaction}) is tedious but standard. We are interested in the low energy dynamics of the effective action so we will keep terms up to order $O(\frac{p^{2}}{\Delta^{2}S^{2}})$ in the expansion parameter $p/\Delta S$. The final result reads\footnote{To compute the loop integrals we used dimensional regularization. In the expression for $\Gamma^{bs}_{||}$ the argument of the logarithm should read $\ln\left( \Delta^{2}S^{2}\theta^{-1}/8\pi\mu^{2}\right)$, where $\mu$ is a fake parameter with dimensions of mass, product of dim. regularization. We fix it to the scale of the low energy physics of the problem, which is $\Delta S$.}:

\begin{subequations}
\begin{eqnarray}
\Gamma^{bs}_{A'}=\frac{1}{4\pi v_F}\int dtdx_2 A'_{a}\left\lbrace\delta_{ab}-\frac{\partial_{a}\partial_{b}}{\partial^{2}}-\frac{i}{2\partial^{2}}(\epsilon_{cb}\partial_{a}\partial_{c}-\epsilon_{ad}\partial_{d}\partial_{b})\right\rbrace A'_{b},\label{effactionfinala}
\end{eqnarray}
\begin{eqnarray}
\Gamma^{bs}_{||}=\frac{\Delta^{2}S^{2}}{4\pi v_F}\int dtdx_2 \left(\gamma+\ln(\frac{\theta^{-1}}{8\pi})\right)\sin^{2}\phi, \label{effactionfinalb}
\end{eqnarray}
\end{subequations}
and $\Gamma^{bs}_{X}=0$. Now we have $\partial_{a}=(\partial_{0},v_F\partial_{2})$, $A'_{a}=(A_{0},v_FA_{2}+\Delta S \cos\phi)$, and remember that $\theta=\Delta S\delta/(2v_F)$. The contribution (\ref{effactionfinala}) is generated by the chiral state alone, and it's gauge variation is not zero, revealing the anomalous nature of the chiral (1+1) theory \cite{JR85} living in the Wall. On the other hand (\ref{effactionfinalb}) is generated by the coupling between the chiral and massive bound states. This term is renormalized by the corresponding term coming from the coupling between the chiral and scattering states, but as we said at the beginning of this section we are not going to compute the former, as the computation would be much more involved. We argue in section \ref{sec:effaction} that this will not affect our final results. Finally the massive-massive coupling does not contribute to this order $O(\frac{p^{2}}{\Delta^{2}S^{2}})$.

We can be more explicit and split $\Gamma^{bs}_{A'}$ in its two contributions coming from the gauge field $A_{a}$ and the part depending on the collective coordinate $\phi(t)$,  $\Gamma^{bs}_{A'}=\Gamma^{bs}_{A}+\Gamma^{bs}_{\phi}$:

\begin{subequations}
\begin{eqnarray}
\Gamma^{bs}_{A}=\frac{1}{4\pi v_F}\int dtdx_2 A_{a}\left(\delta_{ab}-\frac{\partial_{a}\partial_{b}}{\partial^{2}}-\frac{i}{2\partial^{2}}(\epsilon_{cb}\partial_{a}\partial_{c}-\epsilon_{ad}\partial_{d}\partial_{b})\right)A_{b},\label{effactionfinal2a}
\end{eqnarray}
\begin{eqnarray}
\Gamma^{bs}_{\phi}=\frac{\Delta S}{4\pi}\int dtdx\left(2\cos{\phi}\frac{1}{\partial_{0}+i v_F \partial_{2}}E+\frac{\Delta S}{v_F}\cos^{2}\phi-i\frac{1}{v_F}A_{0}\cos\phi\right),\label{effactionfinal2b}
\end{eqnarray}
\end{subequations}
where $E=\partial_{0}A_{2}-\partial_{2}A_{0}$. The terms that violate gauge invariance are the last terms in the integrand of both (\ref{effactionfinal2a}) and (\ref{effactionfinal2b}). Let us stress that the expression (\ref{effactionfinalb}) is valid only at order $O(\frac{p^{2}}{\Delta^{2}S^{2}})$, while expressions (\ref{effactionfinal2a}) and (\ref{effactionfinal2b}) are exact.
\end{widetext}

\section{Contribution of the fermionic scattering states to the effective action for B=0}
\label{app:scatteff}
The propagating states of the fermionic spectrum could potentially contribute to the effective action of the DW, but we will see that their contribution to the (1+1) effective action is going to be zero up to order $O(\frac{p^{2}}{\Delta^{2}S^{2}})$. However, despite having a null (1+1) contribution, the scattering states generate a (2+1) Chern-Simons term that is essential to restore the gauge invariance that was lost in the isolated (1+1) chiral theory \cite{CH85}.

To avoid computing the exact form of the fermionic spectrum we can resort on the adiabatic approximation\cite{UTT11} assuming that the scattering states for a \emph{tanh} type of mass are asymptotically equal to the scattering states for a \emph{constant} mass. This approximation turns to be valid in the low energy limit where $p\ll |\Delta S|$. 
We start from the Hamiltonian for the surface states of the TI and write the fermionic action. As we are now not computing the exact fermionic spectrum (in which case it emerges naturally), it is important to notice that, as the magnetization is translated by $-X(t)$, so have to be the fermions: $\Psi=\Psi(t,x_1+X(t),x_2)$. We have:
\begin{widetext}
\begin{equation}
\mathcal{S}_{TI}=\int dtd^2x\overline{\Psi}(t,x_1,x_2)\Big(\sigma_3(\partial_0+\dot{X}\partial_1)-i\sigma_3A'_0-v_F\sigma_1(\partial_1-iA'_1)-v_F\sigma_2(\partial_2-iA'_2)+m\Big)\Psi(t,x_1,x_2),
\end{equation}
where $A'_0=A_0$, $A'_1=-\Delta S\,sech(x_1/\delta)\sin\phi/v_F$, $A'_2=A_2+\Delta S\,sech(x_1/\delta)\cos\phi/v_F$, $m=-\Delta S\tanh(x_1/\delta)$ (we are taking again a ferromagnetic exchange coupling) and we did a change of variables $x_1\rightarrow x_1-X$. The strategy is to integrate out the fermions with a constant mass $m$, so a Chern-Simons term is obtained, and then substitute $m$ by $-\Delta S\tanh(x_1/\delta)$, arriving to (in imaginary time):
\begin{equation}
\Gamma^{scatt}_{TI}=\frac{i}{8\pi }\int dt d^{2}x sign(x_1)\epsilon_{\mu\rho\nu}A'_{\mu}\partial_{\rho}A'_{\nu}.\label{CSterm}
\end{equation}
\end{widetext}
If we integrate over $x_1$, $\Gamma^{scatt}_{TI}$ vanishes (except a term needed to restore gauge invariance, as we will see in a moment). But the crucial point is that despite it's contribution to the (1+1) effective action being zero, it's gauge variation is finite. If we substitute the explicit value of $A'_\mu$ in (\ref{CSterm}) we get a standard Chern-Simons term for $A_{a}$ (with $a=0,2$), but with a mass that changes sign, and a contribution mixing the components of the gauge field $A_{a}$ and the collective coordinate $\phi$. The gauge variation of the Chern-Simons term cancels exactly the gauge variation of the anomalous chiral (1+1) effective action (\ref{effactionfinal2a}). On the other hand, the mixed term, when integrated over $x_1$, exactly cancels the last term in eq. (\ref{effactionfinal2b}). So the scattering states do not contribute to the effective action of the DW (for $B=0$), but are crucial to restore gauge invariance.

\section{Presence of a perpendicular magnetic field $B$}
\label{app:Bfieldaction}
When an external magnetic field is applied perpendicular to the sample, new contributions to the effective action appear. The first thing to note is that the magnetic field $B$ couples to the fermions both through the gauge field $A_{\mu}$ (orbital contribution) and through a Zeeman term. For small magnetic fields a perturbative treatment of the orbital coupling is justified, but for big fields LL formation becomes relevant and this is not longer the case. However, as explained in the main text, whichever the magnitude of the field the topological response of the (2+1) scattering states is going to be the same.

Regarding the chiral state, it will not see the magnetic field due to it's chiral nature. Then the only contribution will be the chern-simons term generated by the scattering states (it was zero in the case $B=0$, but it will be finite for finite $B$). Of course this is not completely true, because the coupling between the chiral and the massive states could give new contributions also. We can check if this is the case by doing the computation with the bound states, as we did in section \ref{app:boundeff}. To do it, we can treat the orbital coupling perturbatively, as bound states live in (1+1) and the magnetic field is not going to generate a topological non-perturbative response, as is the case in (2+1) with the QHE. Doing the computation one finds that new terms do not appear up to order $O(\frac{p^{2}}{\Delta^{2}S^{2}})$.

Then the only contribution is the topological Chern-Simons term generated by the scattering states. So we can directly import the expression (\ref{CSterm}) and substitute $sign(x_1)$ by $sign(m)$, with $m$ being:
\be
m=-\Delta S\tanh\left(\frac{x_1}{\delta}\right)+\gamma_{TI}B
\ee
For the chiral regime we can write $sign(m)$ as:
\begin{widetext}
\be
sign\left(-\Delta S\tanh\left(\frac{x_1}{\delta}\right)+\gamma_{TI}B\right)=sign\left(-x_1+\delta\tanh^{-1}\left(\frac{\gamma_{TI}B}{\Delta S}\right)\right),
\ee
and for the non-chiral regime:
\be
sign\left(-\Delta S\tanh\left(\frac{x_1}{\delta}\right)+\gamma_{TI}B\right)=sign(B),
\ee
Then, doing the integration in the coordinate $x_1$ on the Chern-Simons term with the $sign$ functions just obtained, we arrive at the following new contributions to the effective action. For the chiral regime:
\be
\Gamma_{TI}^{scatt}=i\int dtdx_2\frac{\delta\gamma_{TI}BE}{2\pi v_F}\sin\phi,
\ee
\end{widetext}
where we did an expansion up to first order in $\gamma_{TI}B/(\Delta S)$. Of course this expansion is justified if $\gamma_{TI}B<< \Delta S$, and this way we can write $\Delta S\tanh(x_1/\delta)+\gamma_{TI}B\approx\Delta S\tanh(x_1/\delta+\gamma_{TI}B/(\Delta S))$ and then linearize the function $\tanh$, so the computation done for $B=0$, based on this linearization, qualitatively holds for $B\neq0$ in the chiral regime. 

On the other hand, for the non-chiral regime we have:
\be
\Gamma_{TI}^{scatt}=iS\int dtdx_2\frac{\Delta\delta sign(B)E}{4v_F}\sin\phi,
\ee

\section{Electromagnetic current configuration in the chiral regime}
\label{app:emcurrent}
Let us go to the chiral regime and calculate the total 1D charge density and charge current density flowing through the wall. From the real time effective action (see main text) we have:
\be
J^{1D}_{0,\pm}=\mp\frac{1}{2\pi}\frac{1}{\partial_0\pm v_F\partial_2}E,
\ee
\be
J^{1D}_{1,\pm}=0,
\ee
\begin{eqnarray}
\nonumber J^{1D}_{2,\pm}&=&\frac{1}{2\pi}\frac{1}{\partial_0\pm v_F\partial_2}\left(v_FE-\Delta S\sin\phi\, \dot{\phi}\right)+\\
&+&\gamma_{eff}\delta B\cos\phi\,\dot{\phi},
\end{eqnarray}
Notice that the current density in the $\hat{\mathbf{x}}_2$ direction has two contributions generated by the motion of the DW, in addition to the electromagnetic contribution which we call $J^{EM}$:
\be
J^{EM}_{0,\pm}=\mp\frac{1}{2\pi}\frac{1}{\partial_0\pm v_F\partial_2}E,
\ee
\be
J^{EM}_{2,\pm}=\frac{v_F}{2\pi}\frac{1}{\partial_0\pm v_F\partial_2}E,
\ee
Due to the anomaly of the 1+1 theory of the Wall, the 1D current density is not conserved:
\be
\partial^\mu J^{1D}_{\mu,\pm}=\mp\frac{E}{2\pi},
\ee

Now we are going to obtain an specific configuration for the electromagnetic current by fixing the boundary conditions. Let us first compute its general form by calculating the integral operator $1/(\partial_0\pm v_F\partial_2)$. We compute $J^{EM}_{2\pm}$, and with the non conservation equation of the 1D current we then obtain $J^{EM}_{0\pm}$. Skipping the explicit calculation we have:
\be
J^{EM}_{0\pm}=\int_tdt'\big[\mp\frac{E}{2\pi}+\partial_2\eta_\pm(t'\mp x_2/v_F)\big]+C_\pm,
\ee
\be
J^{EM}_{2\pm}=\frac{v_F}{2\pi}\int_tdt'E+\eta_\pm(t'\mp x_2/v_F),
\ee
where $\eta_{\pm}$ and $C_\pm$ are a function and a constant, respectively, to be determined by the initial and boundary conditions. We impose $J^{EM}_{2\pm}$ to be time independent and:
\be
J^{EM}_{2\pm}(x_2=-L_2/2)=0,
\ee
so we have:
\be
J^{EM}_{2,\pm}=\pm\frac{E}{2\pi}(x_2+\frac{L_2}{2}).
\ee
If we take an electrostatic field configuration:
\be
A_0(x_2)=-Ex_2+const.
\ee
we can write the current as a function of the voltage between the point $x_2=-L_2/2$ and a given point $x_2$ along the DW:
\be
J^{EM}_{2,\pm}=\mp\frac{1}{2\pi}(A_0(x_2)-A_0(-L_2/2)),
\ee
The averaged current along the DW then is:
\be
I_\pm=\frac{1}{L_2}\int_{-\frac{L_2}{2}}^{\frac{L_2}{2}}dx_2J^{EM}_{2,\pm}=\mp\frac{V}{4\pi},
\ee
with $V=A_0(L_2/2)-A_0(-L_2/2)$ being the voltage between both sides of the magnetic strip. A similar current configuration was recently used in the literature\citep{TL12}.

%\bibliography{DWbiblio}

\end{document}